\documentclass[aps,pre,twocolumn,groupedaddress]{revtex4-2}
\usepackage{amssymb}
\usepackage{amsmath}
\usepackage{color}
\usepackage{sidecap}
\usepackage{graphicx}
\usepackage{bbold}
\usepackage{mathtools}
\usepackage{stackengine}
\usepackage{float}
\usepackage[T1]{fontenc}
\definecolor{MB}{rgb}{0.858, 0.188, 0.478}
\usepackage{xcolor,hyperref}
\hypersetup{
   colorlinks,
   linkcolor={blue!50!black},%{red!80!black},
   citecolor={blue!50!black},
   urlcolor={blue!80!black}
}
\DeclareMathAlphabet{\mathitbf}{OML}{cmm}{b}{it}

\newcommand{\xv}{\mathitbf x}

\newcommand{\Fv}{\mathitbf F}

\newcommand{\calBold}[1]{\mbox{\boldmath${\cal #1}$}}

\definecolor{mycolor}{RGB}{255,0,0} % Red color

\begin{document}

\title{How rigidity percolation and bending stiffness shape colloidal gel elasticity}

\author{David Richard}
\email{david.richard@espci.fr}
\affiliation{PMMH, CNRS, ESPCI Paris, Universit\'e PSL, Sorbonne Universit\'e, Universit\'e Paris Cit\'e, France}
\affiliation{Univ. Grenoble Alpes, CNRS, LIPhy, 38000 Grenoble, France}
\author{Mehdi Bouzid}
\email{mehdi.bouzid@univ-grenoble-alpes.fr}
\affiliation{Univ. Grenoble Alpes, CNRS, Grenoble INP, 3SR, F-38000, Grenoble, France}

\begin{abstract}
Dispersed colloidal particles within a suspension can aggregate and spontaneously self-organize into a robust, percolating structure known as a gel. These network-like structures are prevalent in nature and play a critical role in many industrial processes, including those involving batteries, food products, and pharmaceutical formulations. In this paper, we examine the emergence of elasticity in colloidal gels. We show that gelation is governed by a rigidity percolation transition. We identify a characteristic correlation length that quantifies the extent of elastic and structural inhomogeneities, which diverges at the critical point. Our findings reveal that, regardless of the interaction types, the particle concentration, or the specific route to non-ergodicity i.e. the preparation protocol, the elastic moduli and vibrational properties of gels can be accurately predicted within a unifying framework, in which the bending modes of fractal clusters --approximately the size of this correlation length--  dominate under small deformations.
\end{abstract}

\maketitle

\section{Introduction}
Soft colloidal solids encompass a vast array of familiar substances that we come across daily. These can range from foods and medicines, such as protein gels, to building materials such as cement and advanced polymer composites, as well as slurries. As a result, it is crucial to understand and predict how these materials will respond to various loading scenarios, both from a theoretical and fundamental perspective but also to optimize their mechanical properties. This could entail modifying flow properties for 3D printing applications, creating foods with specific textures, or extending the longevity of colloidal materials used for rechargeable batteries.

Still, while the the elastic and vibrational properties of dense soft systems like granular materials, colloidal glasses or dense emulsions, is now well understood \cite{bonn2017yield,nicolas2018deformation,liu2010jamming}, there is yet no universal scenario for how elasticity arises in soft colloidal gels.

This complexity stems from the intricate and uneven hierarchical structure of colloidal gels \cite{bantawa2023hidden}, where the particles that attract each other within a fluid self-assemble to create a delicate connected network spanning the entire sample. The mechanical behavior of these systems can be adjusted based on the type of interactions between the building blocks, including Van der Waals forces, adhesion, depletion, or electrostatic interactions, among others \cite{lu2008gelation,de2019irreversible,lucey1997formation,muller2025tuning}. Additionally, the volume fraction and the preparation protocol, which govern the gelation process by regulating the temperature and the intensity of interactions, also play a crucial role in determining the resulting non-ergodic gel state \cite{colombo2025kinetic,gao2015microdynamics}. The resulting elastic properties are contingent upon the aforementioned control parameters in a complex and non-trivial manner \cite{gibaud2022nonlinear,bouzid2020mechanics,wang2024distinct}, with their microstructure exhibiting a multiscale and heterogeneous nature \cite{johnson2019influence, richard2018coupling, bouthier2023three, whitaker2019colloidal, mizuno2021structural, nabizadeh2024network}. Several length scales emerge as pertinent for forecasting their elasticity, spanning from the scale of strands, clusters, and pores, which blur the link between the structure and the macroscopic mechanical response.

However, certain prevalent characteristics warrant a universal description: Firstly, although the nature of the aggregation mechanisms may vary across different systems \cite{zaccarelli2007colloidal}, they universally exhibit a transition from a liquid to a solid state \cite{tsurusawa2019direct, fenton2023minimal,zhang2019correlated, rocklin2021elasticity,kohl2016directed,rouwhorst2020nonequilibrium} in which a rigid, interconnected structure extends throughout the entire sample and is referred to as a rigidity percolation transition \cite{jacobs1996generic,zhang2019correlated,javerzat2023evidences}. Secondly, the aggregation of particles results in the development of local resistance to torque, which arises from factors such as the irregular shapes of the particles, the establishment of solid-solid contacts, or the geometric frustration due to the self-assembly of the particles into strands \cite{muller2025tuning,pantina2005elasticity, stuij2019stochastic, bonacci2020contact,bantawa2021microscopic}. In this paper, we present a unifying framework showing that, independently of the nature of the interaction, volume fraction, and preparation protocol, the linear elastic response as well as the vibrational properties of colloidal gels are governed by two emerging properties: i) a local effective bending stiffness at the strand scale and ii) the critical exponent of a rigidity percolation transition. This second-order phase transition marks the emergence of a new critical length scale, similar to that of the jamming transition, which probes the extent of elastic and structural heterogeneities within the network and controls its elasticity.

\section{Rigidity percolation transition }
\begin{figure}[th!]
    \centering
    \includegraphics[width=1\linewidth]{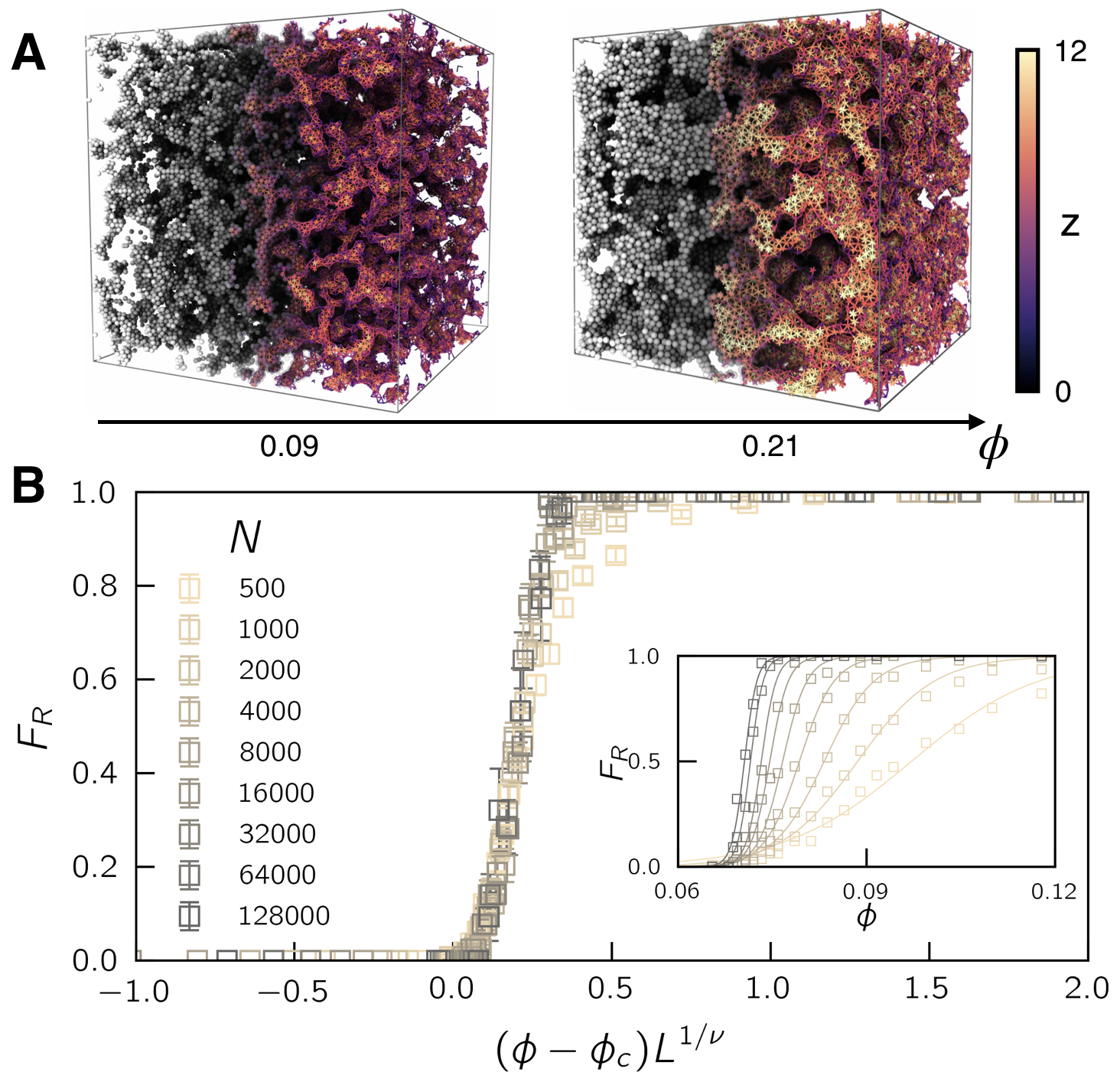}
    \caption{(a) Snapshots of a dilute ($\phi=0.09$) and dense ($\phi=0.21$) gel quenched with $t_q=100$ and $\kappa=0$. The colormap indicates the coordination number $z$. (b) Size collapse of the probability $F_R$ of 3D samples to be rigid using the critical concentration $\phi_c\simeq0.07$ and exponent $\nu\simeq 1$. The inset shows the raw data of $F_R$ versus $\phi$ for different system sizes.}
    \label{fig1:rigidity}
\end{figure}

First, we investigate the critical statistical properties of the transition from a fluid to a solid phase with increasing solid volume fraction $\phi$. We use a standard numerical particle-based model composed of $N$ monodisperse particles that interact via a modified Lennard-Jones short-range attractive potential. Starting from a colloidal fluid, the temperature is quenched for a given time $t_q$, allowing the particles to self-assemble into a percolating, space-spanning network (see details in the SM). Figure.~\ref{fig1:rigidity}(a) shows snapshots of typical configurations where we highlight the change in network connectivity. The overall particle-particle coordination decreases as the volume fraction $\phi$ is reduced, leading to a sparse delicate network with thin stands. Unlike two-dimensional systems, for which a rigorous algorithm for determining rigid clusters from structural analysis is well established~\cite{jacobs1995generic}, three-dimensional systems remain challenging, and no general method has been proposed yet. Furthermore, it is known that a geometrically based counting criterion, such as the Maxwell-Calladine or the pebble game, are not sufficient to describe the rigidity of systems under pre-stress~\cite{huisman2011internal,bose2019self,damavandi2022energetic}. The latter can result from the history of out-of-equilibrium self-assembly~\cite{zhang2022prestressed}, geometrical frustrations~\cite{merkel2018geometrically} or from external mechanical drives~\cite{vermeulen2017geometry,lerner2023scaling}. Since we are dealing exclusively with off-lattice systems that have self-assembled into stressed networks, we therefore rely on mechanical measurements to determine whether the obtained configurations are rigid by systematically measuring the shear $G_0$ and the bulk modulus $K_0$ (see SM for more details).
To quantitatively identify the critical volume fraction $\phi_c$ that separates floppy from rigid gels, we perform a standard finite-size scaling study by generating $n=128-256$ independent configurations for each system size and measure the fraction of rigid samples $F_R$ as a function of $\phi$. As expected for a second-order phase transition, $F_R=F(\phi)$ grows rapidly from $0$ to $1$ at the transition, as shown in the inset of Fig.~\ref{fig1:rigidity}(b). However, we observe strong finite size effects near $\phi_c$, highlighting the presence of a critical point and a corresponding divergent length $\xi$ that has yet to be measured. The data can be collapsed using the following scaling form\cite{zhang2019correlated}:   
\begin{equation}
F_R(\phi,L) = \tilde{F}[(\phi-\phi_c)L^{1/\nu}],
\end{equation}
where $\nu$ is the critical exponent controlling the divergence of $\xi$ according to $\xi\sim \Delta\phi^{-\nu}$, with $\Delta\phi=(\phi-\phi_c)$ . The collapse is shown in Fig.~\ref{fig1:rigidity}(b) with $\phi_c\simeq0.066\pm 0.003$ and $\nu\simeq1\pm0.1$, which agrees -- within the error bar -- with previous results \cite{bantawa2023hidden}. We first note that in our simulations, the limited number of sample realizations and the accessible system size limit our ability to determine the exponent $\nu$ very precisely compared to usual lattice models, yet, and to the best of our knowledge, our statistical analysis constitutes one of the most accurate for three-dimensional off-grid systems. Consequently, for 2D network gels with the same preparation protocol, we observe a similar rigidity transition with a different critical exponent $\nu\simeq1.3\pm0.1$, which falls within the universality class of standard rigidity percolation measured in different lattice model simulations \cite{jacobs1995generic,zhang2019correlated,machlus2021correlated}.

Having identified the rigidity percolation transition, one can now predict the linear elastic response at and far from the critical point. We use a standard mean-field model developed in the context of polymeric systems and recently extended to particulate gels~\cite{bantawa2023hidden}. In this framework, one envisions a percolating backbone formed by thin one-dimensional filaments connecting a set of more rigid regions whose typical separation is of the order of the correlation length $\xi$. In a seminal work, Webman and Kantor have shown that the elastic force constant associated with an elementary string of size $\xi$ can generally be written as $\kappa(\xi)\sim \xi^{-\alpha}$, where $\alpha=1/\nu$ or $\alpha=2+1/\nu$ for a stretching or bending dominated elasticity, respectively \cite{kantor1984elastic}. In analogy to the relation between conductance and resistivity~\cite{de1976relation}, the macroscopic shear modulus $G_0$ and the bulk modulus $K_0$ of the system can be expressed as
\begin{equation}\label{eq:5}
B_0,G_0\sim \frac{\kappa(\xi)}{\xi^{d-2}}\sim(\phi-\phi_c)^\tau\sim\xi^{-\tau/\nu},
\end{equation}
where the exponent $\tau$ reads
\begin{equation}
\tau = \begin{cases}
\nu d +1 -2\nu &\text{stretching}\\
\nu d +1 &\text{bending}
\end{cases}
\end{equation}
We take advantage of the finite size simulations to test this prediction without independently measuring the correlation length~\cite{bergman1985elastic}. As we approach the critical point, the correlation length will grow much larger than the system size. In this regime, one expects $\xi\sim L$, leading to $B_0,G_0 \sim L^{-\tau/\nu}$. This scaling is nicely confirmed in Fig.~\ref{fig:moduli} (SM) for both the shear and bulk modulus in two and three dimensions. The agreement first shows that the stretching modes of the gel are not excited by small deformation unlike the effective bending deformations that control the mechanical response. This is quite remarkable, since no bending rigidity has been introduced in the numerical model, still, while the system self-assembles into a network, the strands has a finite effective bending that dominated over the bond stretching similar to singly connected bonds that develop a finite flexural modulus due to solid-solid contact~\cite{pantina2005elasticity,bonacci2020contact}. This results highlights the importance of strands bending in understanding and modeling the mechanics of colloidal gels \cite{colombo2014stress}.

\section{Measuring elastic heterogeneities}

\begin{figure*}
    \centering
    \includegraphics[width=\linewidth]{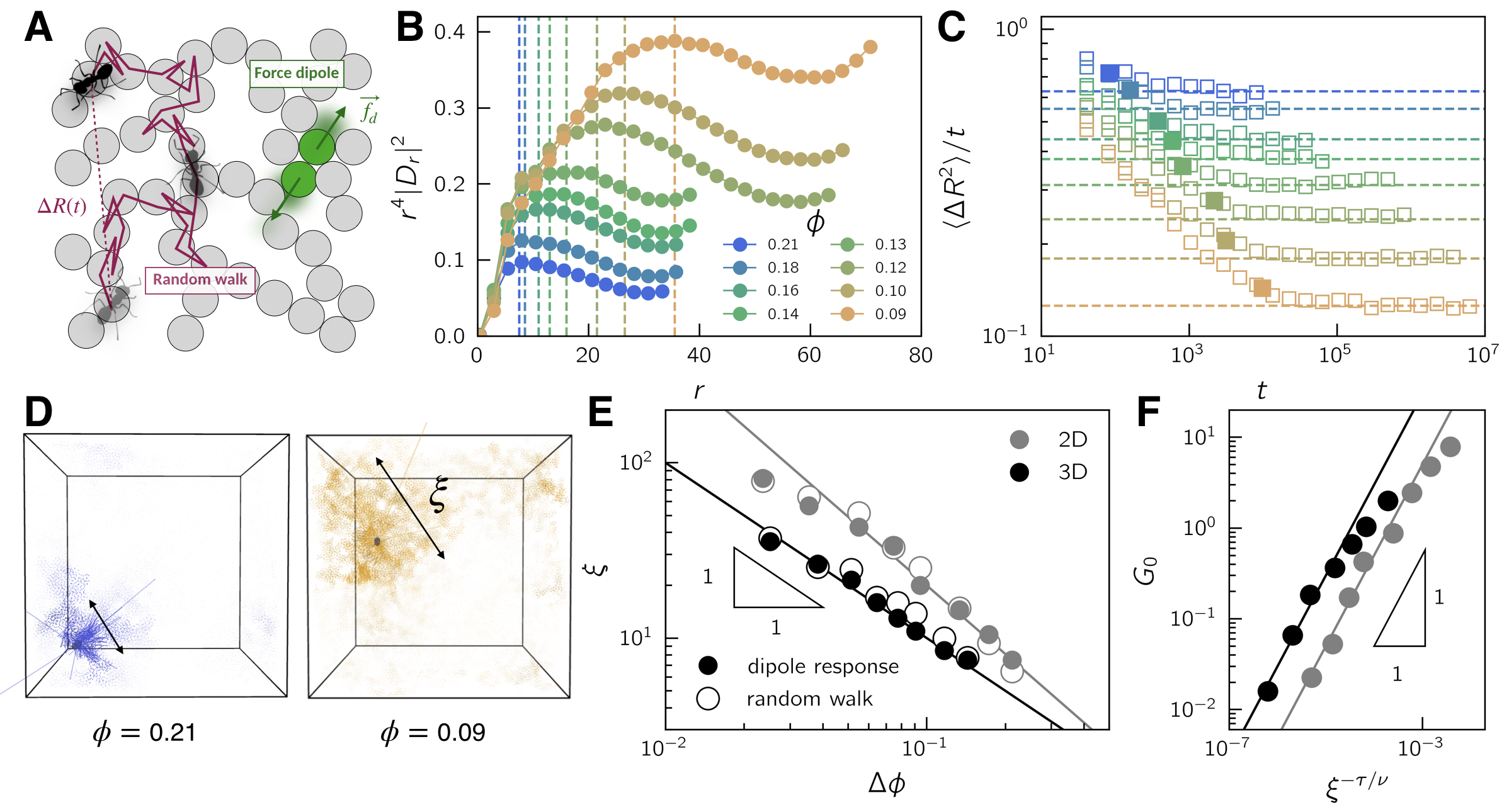}
    \caption{(a) Sketch of an applied dipole force (green arrows) and a tracer ant performing a random walk on backbone of a 2D gel. (b) Linear responses to a local dipole strain normalized by the far field homogeneous elastic decay for various concentrations $\phi$. The vertical dashed lines indicate the length scale $\xi$ above which the system behaves as a continuous homogeneous solid. (c) Mean square displacement of a tracer normalized by the time of the random walk for different $\phi$. The horizontal dashed line indicates the long time and long length scale diffusive behavior. Filled circles indicate the crossover used to extract an estimate for $\xi$. (d) Typical response of a local dipole force in 3D gels for different volume fractions showing the growth of $\xi$ as the  gel approaches the critical point. (e) $\xi$ plotted as a function of the distance to the critical point $\Delta\phi=\phi-\phi_c$ for both 2D and 3D gels. Filled and empty circles correspond to our dipole response and random walk estimates, respectively. (d) Shear modulus $G_0$ plotted against the scaled length $\xi^{-\tau/\nu}$ for both 2D and 3D gels, where $\tau=\nu d+1$. Solid lines show a linear scaling.}
    \label{fig:length_scale}
\end{figure*}

In the thermodynamic limit, this theoretical framework links the linear elastic properties of colloidal gels to the mesoscopic correlation length $\xi$ and the critical exponent $\nu$. However, a major challenge in these multiscale heterogeneous networks~\cite{bouthier2023three} is to determine which structural level of the gel microstructure coincides with the rigidity-percolation correlation length and its physical meaning. A first set of candidates could be a length scale that probes the sparseness and the fractal nature of the network, either through the measure of the integrated radial distribution function, the structure factor and the corresponding length scale that defines the transition from a fractal-like to a homogeneous medium, or by measuring the pore size distribution $P(p_s)$ for different $\phi$ (data shown in the SM), and the average pore size $\langle p_s \rangle$. Our data show that these length scales grow too slowly with the distance to the critical point $\Delta\phi$ compared to the expected critical scaling $\Delta\phi^{-\nu}$. Recently, for particulate gels with explicit bending interaction, it has been shown that the variance of the distribution function of the topological distances connecting the node network is a good proxy for the correlation length~\cite{bantawa2023hidden}. However, extending this approach to colloidal gels formed by densely connected strands and nodes is much more complicated and challenging. Thus, one may wonder how to measure $\xi$ and its physical meaning?

In what follows, we will demonstrate that one can independently measured $\xi$ approaching $\phi_c$ in two ways. At first, we revisit what has been found in the context of the jamming transition in dilute spring networks. Where a spring network of connectivity $z$ is progressively pruned randomly to approach the isostatic point located at $z_c=2d$ \cite{lerner2014breakdown}. Interestingly, despite having no peculiar geometrical network features, it was found that the network exhibits an anomalous elastic response on a length scale $l$ that diverges according to $l\sim 1/\sqrt{z-z_c}$ \cite{lerner2014breakdown}. We test this approach by probing the elastic Green function associated with a dipole force $\vec{f}_d$ applied between two particles that belong to the percolating backbone of our gels, see sketch in Fig.\ref{fig:length_scale}(a). In the linear regime, the response is computed via $\vec{D}_r=\mathcal{H}^{-1}\cdot\vec{f}_d$, with the system Hessian $\mathcal{H}$, see details in the SM. Continuum linear elasticity predicts for the far field decay $|\vec{D}_r|^2\sim r^{2(1-d)}$, with $r$ the distance from the pinched particles. In Fig.\ref{fig:length_scale}(b), we display $r^{4}|\vec{D}_r|^2$ for various concentrations $\phi$ approaching the critical point. We observe a crossover between an anomalous short range scaling and the continuum limit (marked by the vertical dashed line) which is postponed to larger and larger distances as $\Delta\phi\to0$. Examples of dipole responses are shown in 3D snapshots in Fig.\ref{fig:length_scale}(d) highlighting the size of such elastic heterogeneities. The same observation is found in 2D network gels. Plotting this crossover length $\xi$ versus $\Delta\phi$ (Fig.\ref{fig:length_scale}(e)), we can recover the same divergence $\xi\sim(\phi-\phi_c)^{-\nu}$ as previously inferred using finite size scaling, namely $\nu\simeq1.3$ and $\nu=1$ for 2D and 3D gels, respectively. This demonstrate that the same length scale associated with anomalous elasticity as seen in non-fractal jammed packing is at play in controlling the rigidity transition of colloidal gels.

Although monitoring such a response in experiments is technically accessible, e.g., via optical tweezers \cite{jones2015micromechanics}, it can be rather cumbersome. Here, we propose another way to extract $\xi$, knowing only the particle positions within the percolating backbone accessible via confocal imaging \cite{dinsmore2001three}. To probe structural heterogeneities in our system, we borrow concepts developed in percolation theory, namely the anomalous diffusion seen in random walks on fractal objects \cite{rammal1983random,nakayama1994dynamical}. The idea is simple: We place a fictitious tracer particle (illustrated by an ant in Fig.\ref{fig:length_scale}(a)) that is allowed to perform a random walk on the backbone of our networks. At each time step $\Delta t$, the tracer is allowed to jump to a neighboring particle. We monitor the mean square displacement $\langle \Delta R^2 \rangle$ as a function of the walk time $t$. In the long time limit, the tracer simply undergoes normal diffusion where $\langle \Delta R^2 \rangle \sim t$, whereas in the short time limit, one can observe a subdiffusive regime where $\langle \Delta R^2 \rangle\sim t^\beta$, with $\beta<1$, as shown in Fig.~\ref{fig:length_scale}(c) by plotting $\langle \Delta R^2 \rangle/t$ for different concentrations. We locate the crossover between the subdiffusive and diffusive regimes when $\langle \Delta R^2 \rangle/t>a D_\infty$, where $a$ is a parameter set to $1.15$ and $D_\infty$ corresponds to the long time diffusion coefficient, indicated by a horizontal dashed line. This transition can be converted to a length scale that is directly comparable to our previous elastic length scale estimate. Note that the parameter $a$ simply shifts up and down the length scale $\xi$ without affecting its scaling with $\Delta\phi$ as long as $a$ is chosen of the order of unity. Remarkably, we observe that the length scale extracted from the anomalous diffusion coincides perfectly with the scaling of the dipole elastic length. This result highlights that one can get a good proxy for the change in size of the structural heterogeneities controlling the bulk elastic property from the exploration of the fractal network.

At this stage, we have independently measured the critical exponent $\nu$ and the length scale $\xi$. In Fig.\ref{fig:length_scale}(f) we show the shear modulus $G_0$ versus $\xi^{-\tau/\nu}$, with $\tau=\nu d+1$ for both 2D and 3D gels. We observe a remarkable agreement with the equation \ref{eq:5}. Finally, we note that the relationship between $G_0$ and $\xi$ depends on the spatial dimension through $\tau$, which is different from networks close to the jamming transition where one observes $G_0\sim \xi^{-2}$ independently of $d$.

\section{Bending fractons vs. phonons}

Next, we study the vibrational properties of our gels, which remain poorly studied in the literature on off-lattice colloidal gels~\cite{varga2018normal,mizuno2021structural}, and highlight some other differences between the rigidity percolation point and the jamming point of dense packings. Let us first recall that for decompressed sphere packing, one observes in the vibrational density of states (VDoS) $D(\omega)$ an excess population of low-frequency modes that form a plateau ($D(\omega)\sim \omega^0$) above a characteristic frequency $\omega_c$ \cite{liu2010jamming}, below which long wavelength ($\lambda>\xi$) phonons dominate. Upon mechanical driving, this plateau produces large nonaffine motions and a power-law viscoelastic spectrum \cite{baumgarten2017viscous}. Approaching the isostatic point $\Delta z=z-z_c\to 0$, one finds $\omega_c\sim \Delta z$, which follows from the relation $\omega_c\sim c_s/\xi$, with the speed of sound $c_s=\sqrt{G_0/\rho}\sim \Delta z^{1/2}$, with number density $\rho$. This links the crossover frequency to the length scale of the heterogeneities as $\omega_c\sim\xi^2$.

The situation is somewhat different in fractal networks approaching the percolation rigidity transition. First, the mass $M_c$ of the percolating network is also a critical quantity and scales as $M_c\sim(\phi-\phi_c)^{\beta}$. As such, the density $\rho$ entering in $c_s$ cannot be kept constant. Using the hyperscaling relation $\beta=\nu(d-d_f)$ and the exponent from the scaling relation of the shear modulus $\tau = \nu(d+\alpha)-2\nu$, one finds a crossover frequency
\begin{equation}\label{eq:omega_c}
\omega_c\sim\xi^{-(d_f+\alpha)/2},
\end{equation}
which separates phonons into what are often called fractons~\cite{nakayama1994dynamical}. Second, consider a region of size $L$ in our system that is inflated by a factor $b$. The mass and stiffness of the cluster become $M(b L)=b^{d_f}M(L)$ and $\kappa(b L)=b^{-\alpha}\kappa(L)$, respectively. This implies for the frequency $\omega(b L)=b^{-(\alpha+d_f)/2}\omega(L)$, which leads to the well-known scaling for the density of states $D(\omega)\sim \omega^{\tilde{d}-1}$~\cite{alexander1982density,rammal1983random}, where $\tilde{d}$ is the fracton dimension given by~\cite{webman1984elasticity,feng1985crossover,webman1985dynamical}.
\begin{equation}\label{eq:d_tilde}
\tilde{d} = \frac{2 d_f}{\alpha+d_f}= \begin{cases}
2\nu d_f /(1+d_f) &\text{stretching}\\
2\nu d_f / (2\nu+1+d_f) &\text{bending}
\end{cases}.
\end{equation}
In contrast to jammed packings, the low-frequency plateau is not flat and may even slowly diverge when bending elastic modes dominate, as previously shown in lattice models~\cite{webman1985dynamical,nakayama1990density}.

\begin{figure}
    \centering
    \includegraphics[width=\linewidth]{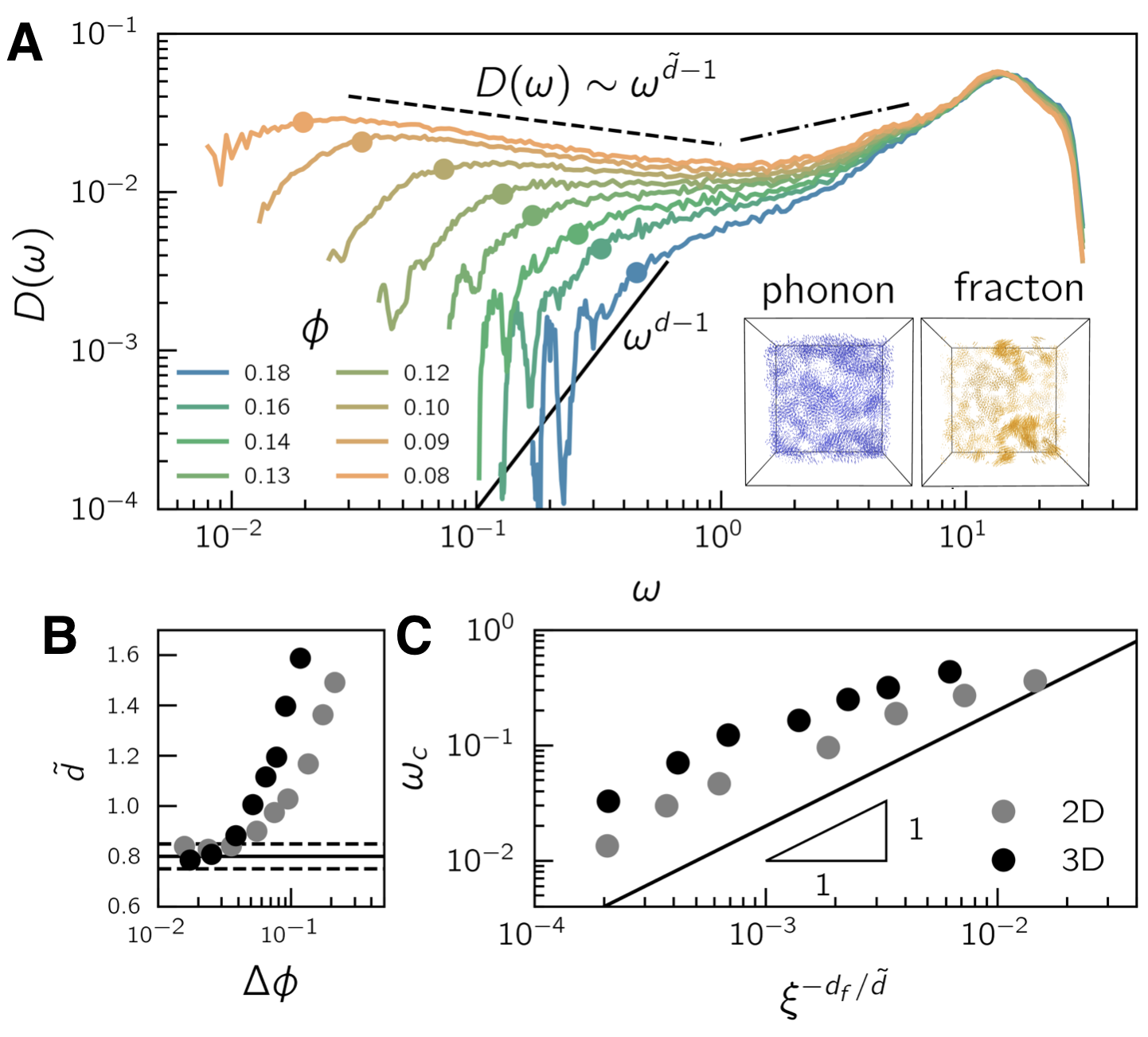}
    \caption{Vibrational density of state $D(\omega)$ for various volume fractions. The solid and dashed lines indicate the Debye $D(\omega)\sim \omega^{d-1}$ and fracton scaling $D(\omega)\sim \omega^{\tilde{d}-1}$, with $d$ and $\tilde{d}$ the spatial and fraction dimension, respectively. The dashed and dotted dashed lines correspond to the scaling of bending and stretching fractons, respectively. The colored circle indicates the crossover between delocalized low-frequency phonons and localized fractons. Modes are displayed in the bottom right corner. (b) Fracton dimension versus the distance to the critical density $\Delta\phi$ for 2D and 3D gels. The horizontal line indicates the asymptotic scaling $\tilde{d}=0.80\pm0.05$. (c) Crossover frequency plotted against the scaled length $\xi^{-d_f/\tilde{d}}$, where $\xi$ is measured from the response of dipole forces.}
    \label{fig:vdos}
\end{figure}

In Fig.\ref{fig:vdos}(a) we plot VDoS for different concentrations approaching $\phi_c$. Below $\omega_c$ we observe discrete bands indicative of transverse plane wave phonons whose frequencies are dictated by the bulk shear modulus and density. For a macroscopic system, these bands form a continuum with a scaling provided by the Debye theory $D(\omega)\sim\omega^{d-1}$. Above $\omega_c$ we observe the presence of a significant population of soft modes described by $D(\omega)\sim\omega^{\tilde{d}-1}$. This phonon-fracton transition is shown in the inset of Fig.\ref{fig:vdos}(a). We find a $\phi$ dependent $\tilde{d}$ that converges to $\tilde{d}\simeq0.8\pm0.05$ approaching $\phi_c$ in both 2D and 3D samples, in very good agreement with the bending prediction described above and lattice calculations~\cite{webman1985dynamical,nakayama1990density}. This is another demonstration that our colloidal gels fall within the Kantor and Webman model in the $\phi\to\phi_c$ limit. Here, blobs of correlation length $\xi$ are composed mainly of thin and tortuous strands that are soft to bending. As $\phi$ increases, the bending fractons will disappear in favor of stretching fractons.

Next, we propose to relate the crossover frequency $\omega_c$ to our length scale $\xi$. Harvesting our previous random walk on percolating backbones, we can infer the fractal dimension $d_f$ by monitoring the volume $V$ explored by our tracer, which should scale as $V\sim R^{d_f}$ (see SM for details). We find $d_f=1.68$ and $d_f=1.9$ in 2D and 3D, respectively. Using Eq.~\ref{eq:omega_c} and Eq.~\ref{eq:d_tilde}, the crossover frequency $\omega_c$ that separates phonons from bending fractons is expected to scale as $\omega_c\sim \xi^{-d_f/\tilde{d}}$. The latter scaling is confirmed in Fig.\ref{fig:vdos}(c) in 2D and 3D gels. This scaling relationship between the crossover frequency and the elastic/structural correlation length scale is a powerful prediction that relates a static quantity to a dynamic time scale that controls the linear viscoelastic response. The associated time scale is expected to capture the transition from fast to slow relaxation modes associated with a terminal time scale, that emphasizes the emergence of a plateau modulus in the viscoelastic spectrum, as well as the large scale dissipation given by the macroscopic viscosity $\eta=G_0/\omega_c$.

\section{Effect of interaction potential and gelation pathway}

\begin{figure}[h!]
    \centering
    \includegraphics[width=\linewidth]{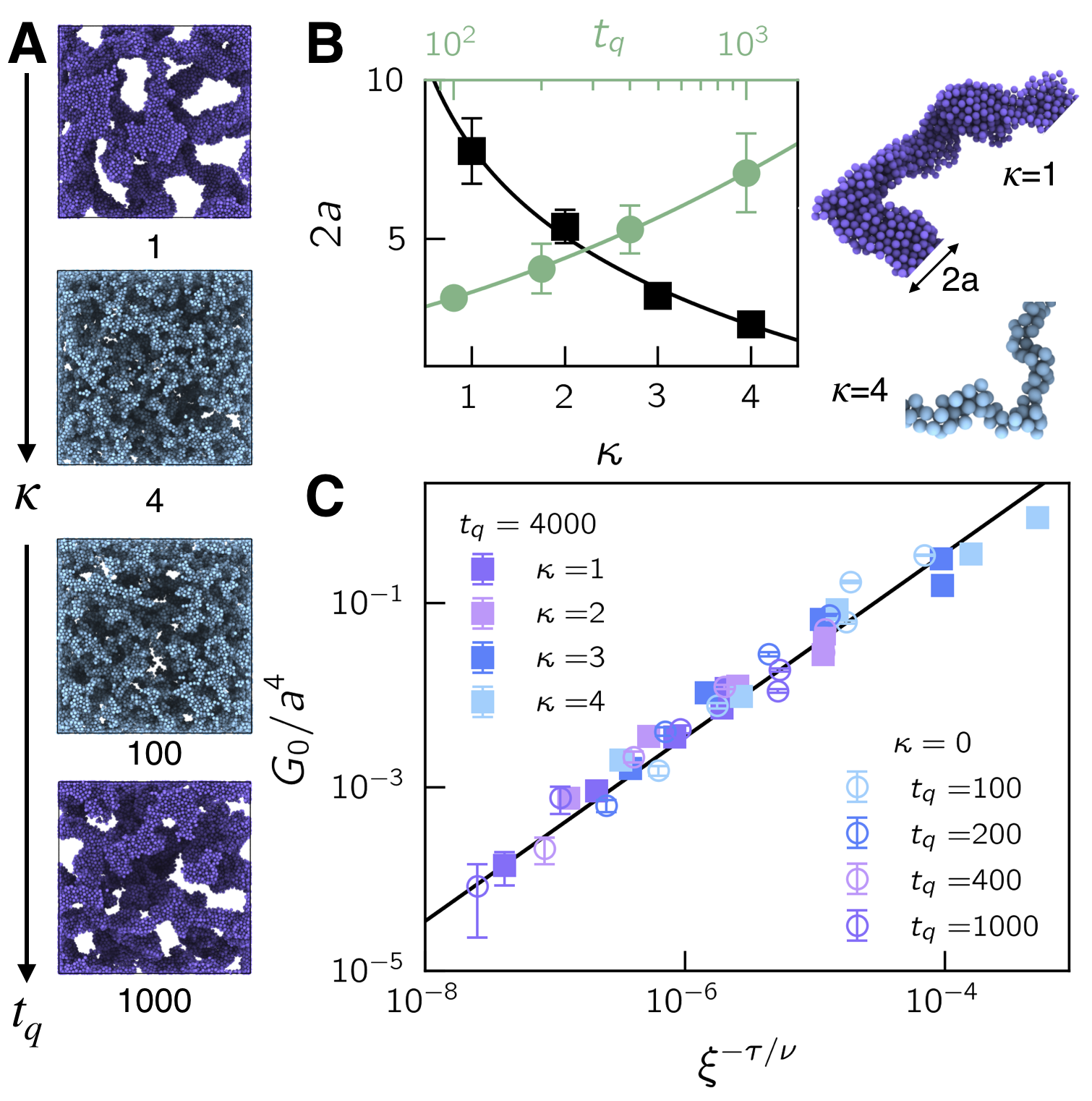}
    \caption{(a) Change of the gel architecture by varying the electrostatic strength $\kappa$ (top) and the quench time $t_q$ (bottom) with $\phi\simeq0.09$. (b) Strand diameter $2a$ plotted as a function of both $\kappa$ and $t_q$. Solid lines are guide to the eye. Snapshots on the right show typical strands for $\kappa=1$ (purple) and $\kappa=4$ (cyan). (c) Collapse of the shear modulus $G_0$ scaled by $a^4$ versus $\xi^{-\tau/\nu}$ in network gels formed by varying $\phi$, $t_q$, and $\kappa$.}
    \label{fig:strand_size}
\end{figure}

Having characterized the critical behavior in the VDoS of colloidal gels and firmly established the main contribution of bending modes to the elastic response. We now examine how the linear elastic properties are affected by i) the addition of a long-range Yukawa repulsion that accounts for electrostatic contributions, where its strength is controlled by $\kappa$ (See SM) and, ii) the pathway through which gelation occurs given by the quenching time $t_q$. In Fig.~\ref{fig:strand_size}(a) we show both the effect of $t_q$ and $\kappa$ on the microstructure of a gel. We start with a fixed quench time ($t_q=4000$) and gradually increase $\kappa$. For small $\kappa$, the kinetics of the aggregation lead to gel formed by arrested phase separation and form large colloidal strands. Increasing $\kappa$ prevents the coarsening due to medium range repulsion. Without electrostatic $\kappa=0$, increasing $t_q$ leads to coarser and coarser gels. We quantify the effect of both varying $\kappa$ or $t_q$ by extracting the typical strand diameter $2a$, see the results in Fig.~\ref{fig:strand_size}(b) and strand snapshots for illustrations. We observe a variation of the strand radius by a factor of $3$. At first glance, this change may seem insignificant, but in fact it has a profound effect on the macroscopic elasticity of the gel. For a beam of radius $a$, considered as an isotropic solid with a young modulus $E$, the bending modulus can be expressed as follows
\begin{equation}
 k_0=\frac{\pi}{4}Ea^4.
\end{equation}
Since $k_0$ is the prefactor of the effective stiffness $\kappa(\xi)$ of singly bonded strands, the bulk elastic properties can change by more than an order of magnitude. This is confirmed by comparing the shear modulus $G_0$ of the sample shown in Fig.~\ref{fig:strand_size}(a) with $G_0\simeq 0.032$ and $G_0\simeq 0.0035$ for $\kappa=1$ and $\kappa=4$, respectively. Finally, we propose to take our entire gel catalog with different concentrations $\phi$, quench time $t_q$, and repulsion strength $\kappa$ and plot $G_0/a^4$ versus $\xi^{\tau/\nu}$, where $\xi$ is extracted from our random walk, see Fig.~\ref{fig:strand_size}(c). We observe a master curve of unity slope for our entire gels library. These results confirm that, regardless of the nature of the interaction, the volume fraction, or the protocol of aggregation, the elastic properties of the resulting networks are controlled by the critical nature of the rigidity percolation transition, which is associated with a unique critical exponent $\nu$, and the corresponding size of the structural and elastic heterogeneities that build up upon solidification.

\section{Summary and Outlook}

Our work provide a universal framework that rationalize the elastic response of colloidal gels, by firmly establishing the role that rigidity percolation transition plays in controlling a diverging elastic length scale $\xi$, as well as the self-organization of finite colloidal strands giving rise to a local effective bending stiffness. The interplay between sample density, the nature of the interactions and gelation time results in different network architectures that all fall within the same universal master curve $G_0/a^4\sim \xi^{-\tau/\nu}$. Our results also suggest that $\xi$ can be extracted based on the sole knowledge of the particle network, hence, could be measured in colloidal experiments using confocal microscopy~\cite{dong2022direct}.

Furthermore, we have observed that below the correlation length $\xi$ the response to a local mechanical perturbation decays much weaker compared to the predicted continuum linear elasticity, such as seen in many different disordered systems close to a marginal point from fiber networks~\cite{mao2011criticality,sharma2016strain,head2005mechanical,goren2023probing}, dense suspension flows~\cite{during2014length} to nearly jammed amorphous solids~\cite{lerner2014breakdown,lerner2023anomalous}. This anomalous behavior stems from the presence of soft quazilocalized modes above the critical frequency  $\omega_c$, which become increasingly delocalized near the rigidity transition. However, in contrast to canonical homogeneous jamming physics systems, colloidal gels dominated by bending modes exhibit different low-frequency statistics dedicated by the spectral dimension $\tilde{d}$. How the fractal dimension $d_f$ and the critical exponent $\nu$ entering in $\tilde{d}$ shape the scaling of this anomalous elastic behavior will require further investigations. 

We also expect that the viscoelastic spectrum can be derived analytically from the vibrational density of states~\cite{lemaitre2006sum,baumgarten2017viscous} in which the population of bending fractons and their magnitude ($D(\omega)\sim\omega^{\tilde{d}-1}$) control the exponent of a power law regime at at intermediate frequencies ~\cite{bouzid2018computing,bantawa2023hidden}, while $\omega_c$ dictate the appearance of a frequency-independent plateau shear modulus in the limit of $\omega\to 0$.  

Finally, the present work focuses on centro-symmetric pairwise interactions and the conventional gelation protocol under quiescent conditions. The effect of particle roughness~\cite{wang2019surface,bonacci2020contact} and shape~\cite{burger2025preparation}, as well as colloidal gels formed under mechanical loads such as flow-induced aggregation or shear history~\cite{colombo2025kinetic,moghimi2017colloidal} requires further investigation. How the universality class of the rigidity percolation transition --i.e., its critical exponents that have a pivotal role in inferring the elastic response-- is affected by these more complex scenarios thus remains an open question.

{\em Acknowledgments.} M.B. and D.R. thank E. Del Gado, T. Divoux, T. Gibaud, S. Manneville, E. Lerner, P. Lettinga and H. Herrmann for insightful discussions, and M. Mugnai for sharing codes to compute the pore size.  D.R.~acknowledges support by the H2020-MSCA-IF-2020 project ToughMG  (No.~101024057) and computational resources provided by GENCI (No.~AD010913428). M.B.~acknowledges support by the ANR grant ANR-22-CE06-0036-03.

%\bibliography{references}


% \cleardoublepage

\onecolumngrid
\newpage
\begin{center}
	\textbf{\large Supplementary material: ``How rigidity percolation and bending stiffness shape colloidal gel elasticity''}
\end{center}

\setcounter{figure}{0}
\renewcommand{\figurename}{Fig.}
\renewcommand{\thefigure}{S\arabic{figure}}

The goal of this document is to provide additional technical details regarding the results presented in the main text and some supporting results.

\section{Model and numerical details}

\subsection{Model and sample preparation}

We use a coarse-grained molecular dynamics simulations of a model colloidal gel with. Our system is composed of a monodisperse assembly of $N$ particles with density $\rho=N/V$, with $V$ the volume of our simulation box. The volume fraction (concentration) is computed as $\phi=\pi\rho/4$ and $\phi=\pi\rho/6$ in 2D and 3D, respectively. We define particle coordinates and forces via the vector $\xv$ and $\Fv$, respectively. During gel formation, particles of equal mass $m$ interact with a modified Lennard-Jones potential~\cite{dauchot2011athermal} that reads
\begin{equation}
    \varphi_S(r) \!=\!
\left\{
\begin{array}{cc}
\!\!4\epsilon \bigg[ \big(\frac{\sigma}{r}\big)^{12} - \big(\frac{\sigma}{r}\big) ^{6} \bigg],
     &  \frac{r}{\sigma} <  x_{\mbox{\tiny min}}  \\
\epsilon \bigg[a\big(\frac{\sigma}{r}\big)^{12} -b\big(\frac{\sigma}{r}\big)^{6} + \sum\limits_{\ell=0} ^{3}  c_{\mbox{\tiny $2\ell$}} \big(\frac{r}{\sigma}\big)^{2\ell} \bigg] , & x_{\mbox{\tiny min}}\!\le\! \frac{r}{\sigma}< x_c\\
0\,,  & x_c \le \frac{r}{\sigma}
\end{array}
\right. ,
 \label{eq:potential}
\end{equation}
where $\epsilon$ is a microscopic energy scale, $x_{\mbox{\tiny min}},x_c$ are the (dimensionless) locations of the minimum of the Lennard-Jones potential and modified cutoff, respectively, and the $\sigma$ is the length parameters. We express the dimensionless cutoff $x_c$ in terms of $x_{\mbox{\tiny min}}\!=\!2^{1/6}\sigma$, for simplicity, by defining $r_c\!\equiv\!x_c/x_{\mbox{\tiny min}}$. In this work $r_c$ is fixed and equal to $1.2\sigma$. 

We also consider gelation with competing interactions, where we introduce, in addition to $\varphi_{S}$, a long-ranged repulsive Yukawa potential that accounts for electrostatic contributions:
\begin{equation}
    \varphi_{Y}(r) \!=\! \kappa  e^{-r/l_s},
\end{equation}
with varying amplitude strength $\kappa$, fixed Debye screening length $l_s=1/3 \sigma$, and a cutoff set at $r=3\sigma$ beyond which $\varphi_{Y}=0$. Increasing the strength $\kappa$ allows one to postpone structural coarsening during a finite quench. Throughout this study, quantities are expressed in simulation units, namely lengths, energies, and time in units of $\sigma$, $\epsilon$, and $\sqrt{m\sigma^2/\epsilon}$.

We first equilibrate our system in the NVT ensemble using conventional Molecular Dynamics (MD) with a Nose-Hoover thermostat. We set the equilibration temperature to $k_BT/\epsilon=2$, which is much larger than the location of the spinodal and onset of phase separation. Subsequently, we quench our system to a low temperature $k_BT/\epsilon=0.01$ within a quench time $t_{\rm quench}$. We then deactivate the Yukawa potential and only keep the short-range attractive $\varphi$ potential. The system is then brought to mechanical equilibrium ($\Fv=0$) using an energy minimization. Our gel catalog is obtained by varying three control parameters, namely: (i) the particle concentration $\phi$, (ii) the quench time $t_{\rm quench}$, and (iii) the strength of repulsion $\kappa$.

\subsection{Shear and bulk moduli}

The potential energy of our gels is given by $U\!=\!\sum_{i<j}\varphi_{S}(r_{ij})$, where the pairwise potential we employed is described in Eq.~(\ref{eq:potential}). As we focus on athermal conditions, the zero-frequency shear modulus is defined as~\cite{lutsko1989generalized}
\begin{equation}
    G_0 \equiv \frac{1}{V}\left(\frac{\partial^2U}{\partial\gamma^2} - \frac{\partial^2U}{\partial\gamma\partial\xv}\cdot\calBold{H}^{-1}\cdot\frac{\partial^2U}{\partial\xv\partial\gamma} \right)\,,
\end{equation}
where $\calBold{H}\equiv\frac{\partial^2U}{\partial\xv\partial\xv}$ is the Hessian matrix of the potential $U$. Strain derivatives follow from coordinates transform via $\xv\!\to\!\bm{D}(\gamma)\cdot\xv$ with the parameterized deformation tensor (in 3D)
\begin{equation}\label{shear_transformation_matrix}
\bm{D}(\gamma) =  \left( \begin{array}{ccc}1&\gamma&0\\0&1&0\\
0&0&1\end{array}\right)\,.
\end{equation}
The pressure is given by $p\equiv-\frac{1}{Vd}\frac{\partial U}{\partial \eta}\,$, where $d$ is the dimension of space, and $\eta$ is the isotropic dilatational strain that parametrizes the suitable transformation of coordinates $\xv\!\to\!\bm{D}(\eta)\cdot\xv$ as (in 3D)
\begin{equation}\label{dilation_transformation_matrix}
\bm{D}(\eta) =  \left( \begin{array}{ccc}e^\eta&0&0\\0&e^\eta&0\\0&0&e^\eta\end{array}\right)\,.
\end{equation}
The athermal bulk modulus $K_0\!\equiv -\frac{1}{d}\frac{dp}{d\eta}$ reads
\begin{equation}\label{eq-K}
    K_0= \frac{1}{Vd^2}\left(\frac{\partial^{2}U}{\partial \eta^{2}} -d \frac{\partial U}{\partial\eta}- \frac{\partial^{2}U}{\partial \eta \partial \xv} \cdot \calBold{H}^{-1} \cdot \frac{\partial^{2}U}{\partial\xv\partial\eta}\right)\,.
\end{equation}

\subsection{Vibrational Density of States (VDoS)}

The VDoS of small computer gels has been measured via a complete diagonalization of the Hessian matrix $\calBold{H}$. We denote the kth eigenvector by $\calBold{\psi}^k$ with mode frequency $\omega^k$. We compute its spatial extension from the participation ratio defined as
\begin{equation}
e_p^k=\frac{\Big(\sum_i \calBold{\psi}^k_i\cdot \calBold{\psi}^k_i\Big)^2}{N\sum_i(\calBold{\psi}^k_i\cdot \calBold{\psi}^k_i)^2},
\end{equation}
where $\calBold{\psi}^k_i$ denotes the cartesian vector of the ith particle. Finally, we compute the coupling between each eigenvector and the shear force $\Theta=- \frac{\partial^2U}{\partial\gamma\partial\xv}$. The latter allows to evaluate the mode contribution to the non-affine part of the shear modulus~\cite{lemaitre2006sum}
\begin{equation}
\mathcal{I}_{\rm naf}(\omega) = \frac{D(\omega)\langle (\Theta\cdot\psi)^2 \rangle_\omega}{\omega^2},
\end{equation}
where $\langle ... \rangle_\omega$ has to be understood as a running average over modes with a frequency $\omega+d\omega$.

The VDoS of large computer gels was obtained using the Kernel Polynomial Method (KPM)~\cite{weisse2006kernel}, following the procedure described in great detail in Ref.~\cite{beltukov2016boson}. This method relies on two parameters, namely the truncation degree $K$ of the polynomial series and the number $R$ of initial random vectors. Throughout this study, we use $K=100000$ and $R=10$. Furthermore, the VDoS is averaged over $32$ independent gels for each state point.

\subsection{Linear response to dipole forces}

Having the Hessian of the system at hand, one can probe local elastic heterogeneities inside the gel by computing the linear response $\vec{u}_\alpha$ to a dipole force $\vec{d}_\alpha$ applied on a pair of particles $\alpha=\{i,j\}$ as~\cite{lerner2014breakdown}
\begin{equation}
\calBold{H}\cdot \vec{u}_\alpha = \vec{d}_\alpha.
\end{equation}
Here we have chosen random pairs of particles belonging to the percolating network with coordination number $z\ge2d$, where $d$ is the spatial dimension. For each concentration we have measured $\vec{u}_\alpha$ for about $3000-6000$ independent pairs. Finally, we compute the radial profile of the square norm $|D_r|^2$, which is measured as the running median over all $|\vec{u}_\alpha|^2$.

\subsection{Rigidity detection}

\begin{figure*}[h!]
   \includegraphics[width = 1\textwidth]{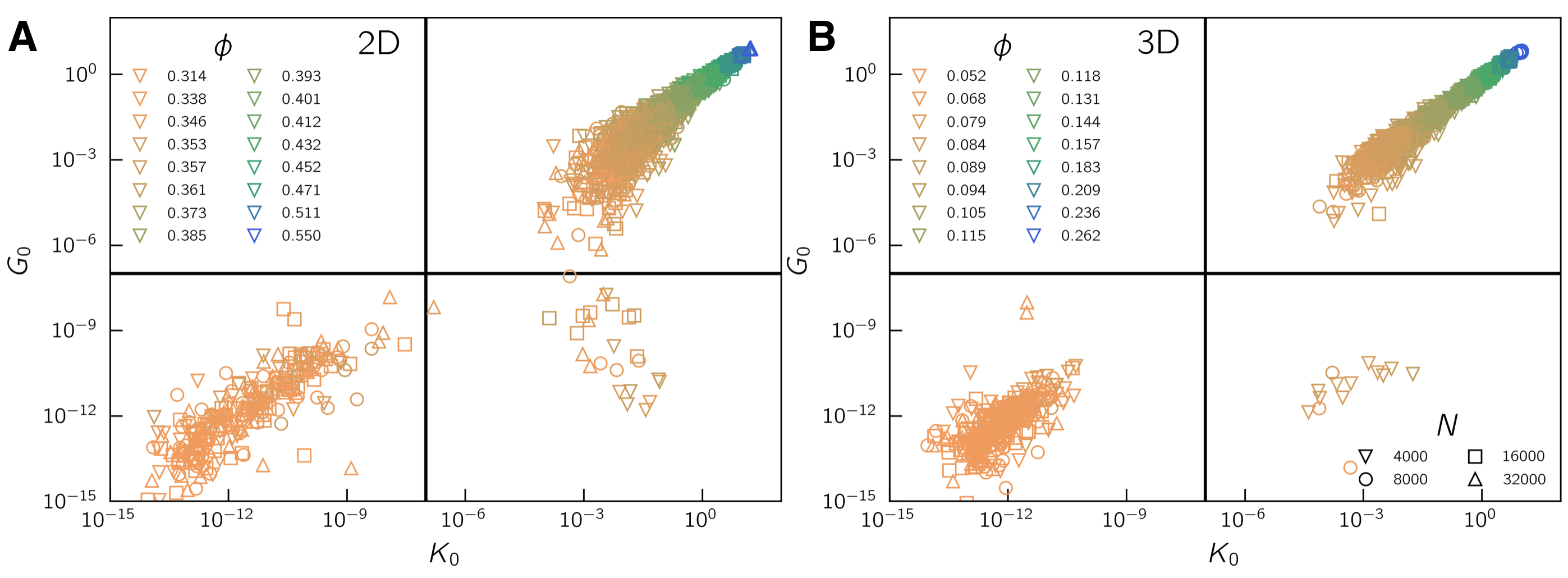}
  \caption{\footnotesize (a) 2D zero frequency shear modulus $G_0$ plotted against bulk modulus $K_0$ for different samples of different size $N$ (different symbols) prepared at different concentrations $\phi$ (different colours). The vertical and horizontal lines indicate the criteria used to discriminate between floppy and rigid networks with $G_0>10^{-7}$ and $K_0>10^{-7}$ respectively. (b) Same as (a) but for 3D samples. In both panels gels are prepared with $t_q=100$ and $\kappa=0$.}
  \label{fig:rigidity_detection}
\end{figure*}

In this work, we consider both 2D and 3D network gels, so we cannot rely on the pebble game introduced by Jacobs and Thorpe to identify floppy from rigid samples \cite{jacobs1995generic}. Here we use the analytical expressions for both the zero frequency shear modulus $G_0$ and the bulk modulus $K_0$ (described above) to determine whether the percolating backbone has finite elasticity. As shown in Fig.~\ref{fig:rigidity_detection}(a) and (b) for 2D and 3D samples respectively, we can reliably label a sample as rigid using the criteria $G_0>10^{-7}$ and $K_0>10^{-7}$. The same approach was used for vertex models to determine the stiffness of confluent tissues~\cite{merkel2018geometrically}.

%%%%%%%%%%%%%%%%%%%%%%%%%%%%%%%%%%%%%%%%%%%%%%%%%%%%%%%%%%%%%%%%%%%%%%%
%%%%%%%%%%%%%%%%%%%%%%%%%%%%%%%%%%%%%%%%%%%%%%%%%%%%%%%%%%%%%%%%%%%%%%%

\subsection{Rigidity fractions and exponents}

\begin{figure*}[h!]
   \includegraphics[width = 1\textwidth]{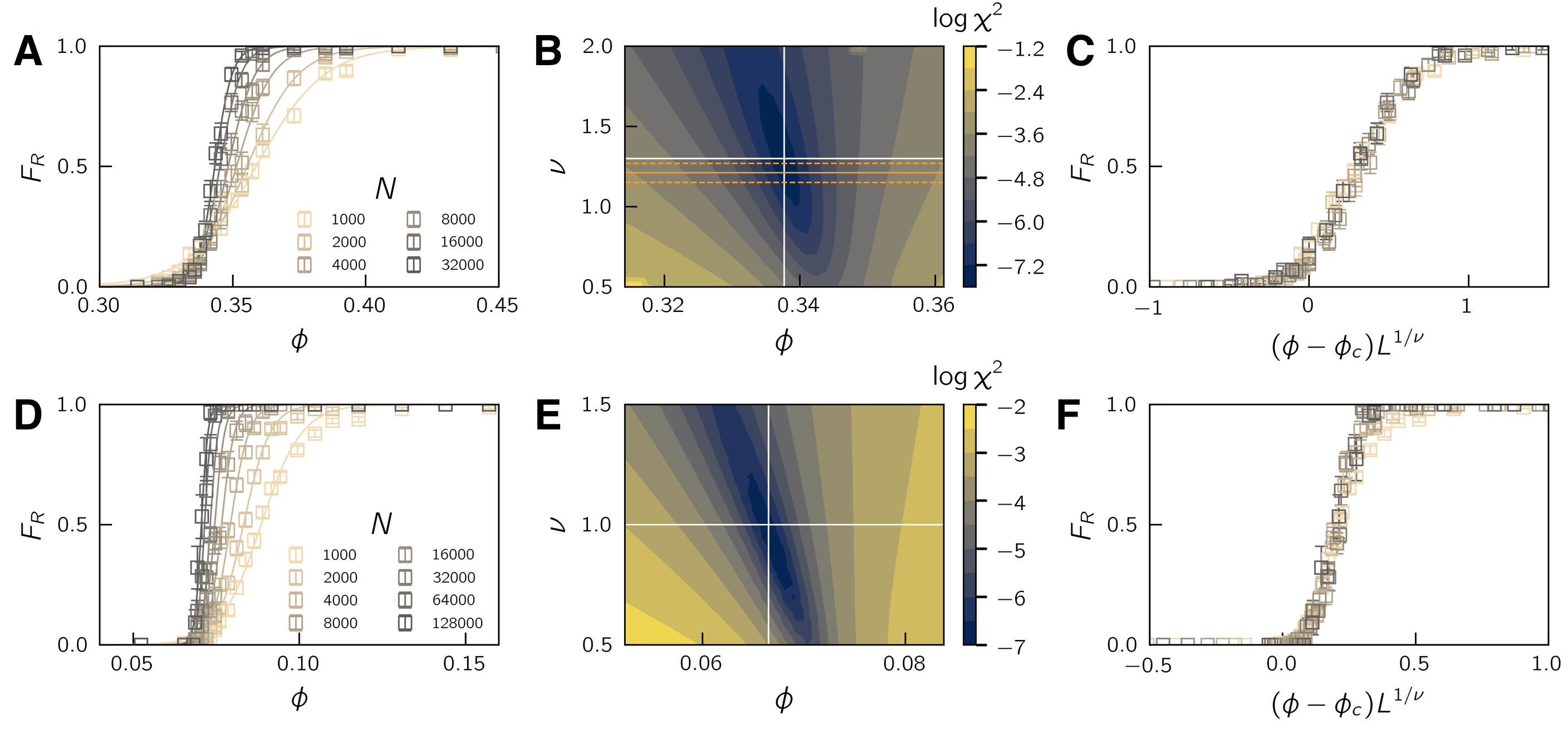}
  \caption{\footnotesize (a) Fraction of rigid sample $F_R$ as a function of the volume fraction $\phi$ for different system sizes in 2D. Solid lines are fit according to Eq.~\ref{eq:profile}. (b) Colormap of the collapse error in the plane ($\phi-\nu$). The orange horizontal line indicates the estimation range of the rigidity transition $\nu=1.21(6)$. White vertical and horizontal lines indicate our estimate for $\phi_c$ and $\nu$, respectively. (c) Size collapse of the rigidity fraction using $\phi_c\simeq0.34$ and $\nu\simeq1.3$. (d), (e), and (f) are the same as (a), (b), and (c) but for 3D gels. The size collapse is obtained using $\phi_c\simeq0.066$ and $\nu\simeq1$. Gels are formed using $A=0$ and $t_{\rm quench}=100$.}
  \label{fig:rigidity}
\end{figure*}

In this section, we provide supporting information on the extraction of the critical volume fraction $\phi_c$ and in exponent $\nu$ in 2D and 3D network gels using $A=0$ and $t_{\rm quench}=100$. In fig.~\ref{fig:rigidity}(a), we show the fraction of rigid sample $F_R$ as a function of the concentration $\phi$ for different system sizes. We define a sample as rigid if $G_0>10^{-6}$ and $K_0>10^{-6}$. The fraction is well modeled by
\begin{equation}\label{eq:profile}
F_R(\phi)=\frac{1}{2}\Big[1+\tan{\Big(\frac{\phi-\phi_0}{\Delta}\Big)}\Big],
\end{equation}
where $\phi_0$ and $\Delta$ are the density crossover and extent, respectively. Following Ref.~\cite{zhang2019correlated}, we expect the scaling form $F_R(\phi,L)=\tilde{F}[(\phi-\phi_c)L^{1/\nu}]$. Here, we explore the plane $\phi-\nu$, where for each parameter set, we plot $F_R(\phi,L)$ versus $(\phi-\phi_c)L^{1/\nu}$ and fit it according to Eq.~\ref{eq:profile}, giving us an estimate function $F_{\rm e}$. We then compute the collapse error as 
\begin{equation}\label{eq:chi}
\chi^2= \frac{1}{n_p}\sum_p (F_R^p-F_{\rm e}^p)^2,
\end{equation}
where $F_R^p$ is the fraction of data point $p$ with concentration $\phi_p$ and system size $L_p$, $F_{\rm e}$ is the fit function evaluated at $x=(\phi_p-\phi_c)L_p^{1/\nu}$, and $n_p$ is the number of points in our data set. $\chi^2$ is the measure of the spread of our data compared to $F_{\rm e}$. In fig.~\ref{fig:rigidity}(b) we show the logarithm of $\chi^2$ in the plane $\phi-\nu$. The white lines indicate the minimum of $\chi^2$ and our estimate for $\phi_c\simeq0.34$ and $\nu\simeq1.3$. The resulting finite-size collapse is shown in fig.~\ref{fig:rigidity}(c). However, we stress that our statistics and system sizes are rather limited in off-lattice systems, so our exponent $\nu\simeq1.3$ is only a rough estimate. Nevertheless, it is close to the 2D rigidity percolation exponent $\nu\simeq1.21(6)$ \cite{jacobs1995generic} and is also consistent with the divergence of length $\xi$ plotted against $\phi-\phi_c$ (see main text). The same analysis for three-dimensional gel networks is shown in Fig.~\ref{fig:rigidity}(d-f).

%%%%%%%%%%%%%%%%%%%%%%%%%%%%%%%%%%%%%%%%%%%%%%%%%%%%%%%%%%%%%%%%%%%%%%%
%%%%%%%%%%%%%%%%%%%%%%%%%%%%%%%%%%%%%%%%%%%%%%%%%%%%%%%%%%%%%%%%%%%%%%%

\subsection{Shear and bulk modulus finite size study}

\begin{figure*}[h!]
   \includegraphics[width = 1\textwidth]{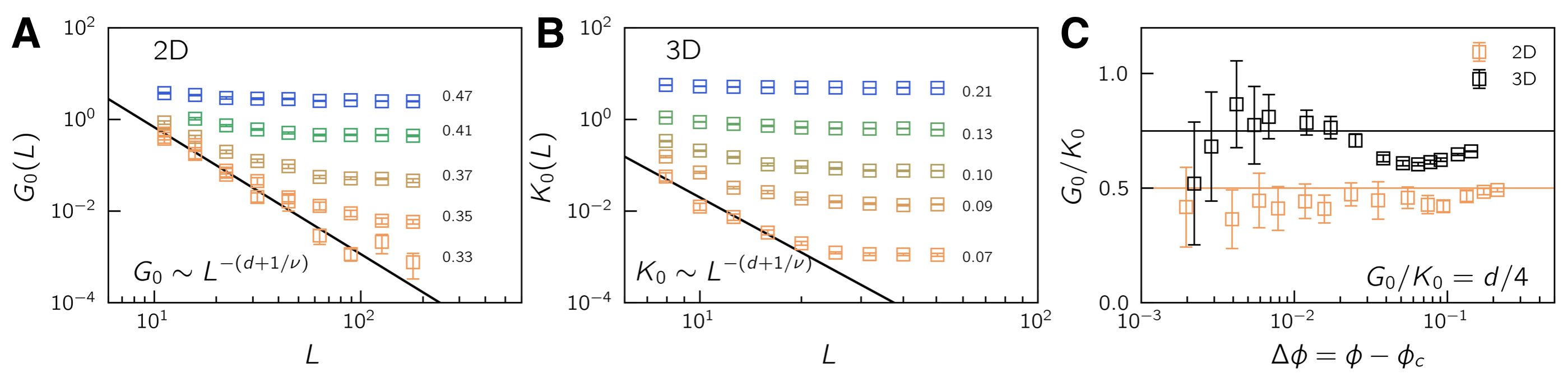}
  \caption{\footnotesize (a) Shear modulus $G_0$ as a function of box length $L$ for different volume fractions in 2D gels. The solid black line indicates the scaling $G_0\sim L^{-(d+1/\nu)}$ with $\nu=1.3$. (b) Bulk modulus $K_0$ as a function of box length $L$ for different volume fractions in 3D gels. The solid black line indicates the scaling $K_0\sim L^{-(d+1/\nu)}$ with $\nu=1$. (c) The ratio $G_0/K_0$ as a function of the distance to the rigidity point $\Delta\phi=\phi-\phi_c$ for 2D (orange) and 3D (black) gels. The solid lines correspond to $G_0/K_0=d/4$. Gels are formed with $A=0$ and $t_{\rm quench}=100$.}
  \label{fig:moduli}
\end{figure*}

In the vicinity of the critical point we expect the shear and bulk moduli to scale as $G_0,K_0\sim (\phi-\phi_c)^{\tau}$, and therefore alternatively we can express the bulk elastic properties as a function of correlation length as $G_0,K_0\sim \xi^{-\tau/\nu}$. However, in a finite system $\xi\to L$ approaching $\phi_c$ we expect size dependent elastic moduli $G_0,K_0\sim L^{-\tau/\nu}$. We confirm such finite size effects with 2D shear moduli in Fig.~\ref{fig:moduli}(a) and 3D bulk moduli in Fig.~\ref{fig:moduli}(b). Close to $\phi_c$ we find that the data are well modelled using the Kantor and Webman exponent $\tau=\nu d +1$ \cite{kantor1984elastic}, giving $G_0,K_0\sim L^{-(d+1/\nu)}$. This result confirms that our system falls within the universality of the bending dominated regime, despite the absence of explicit local bending as previously introduced in lattice models~\cite{feng1984percolation,arbabi1988elastic}. In addition, in solvable fractal models of percolating networks, the shear to bulk modulus ratio was shown to follow $G_0/K_0=d/4$~\cite{bergman1984critical}. In Fig.~\ref{fig:moduli}(c) we confirm this result in our off-lattice gels by plotting $G_0/K_0$ versus the distance to the rigidity point $\Delta \phi = \phi-\phi_c$ for 2D and 3D samples.

%%%%%%%%%%%%%%%%%%%%%%%%%%%%%%%%%%%%%%%%%%%%%%%%%%%%%%%%%%%%%%%%%%%%%%%
%%%%%%%%%%%%%%%%%%%%%%%%%%%%%%%%%%%%%%%%%%%%%%%%%%%%%%%%%%%%%%%%%%%%%%%

\subsection{Dipole responses and random walk in two-dimensional gels}

\begin{figure*}[h!]
   \includegraphics[width = 1\textwidth]{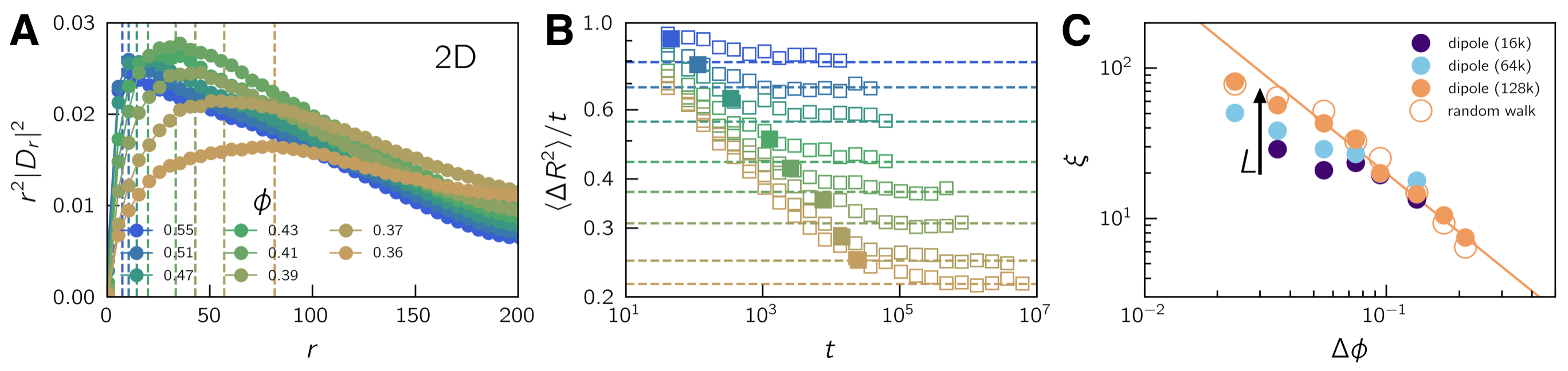}
  \caption{\footnotesize (a) 2D linear responses to local dipole strain normalised by far-field homogeneous elastic decay for different concentrations $\phi$ with $N=128k$. The vertical dashed lines indicate the length scale $\xi$. (b) Mean square displacement of the random walk normalised by time for different $\phi$. The horizontal dashed line indicates the long time and long length scale diffusive behaviour. Filled circles indicate the crossover used to extract an estimate for $\xi$. (c) $\xi$ plotted as a function of distance to critical point $\Delta\phi=\phi-\phi_c$. The solid line indicates the scaling $\xi\sim\Delta\phi^{-1.3}$. The arrow indicates finite size effects approaching $\phi_c$. Gels are formed with $A=0$ and $t_{\rm quench}=100$.}
  \label{fig:length}
\end{figure*}

In this section we provide additional data for the extraction of the length scale $\xi$ in two-dimensional gels. In Fig.~\ref{fig:length}(a) we show the square of the norm of the dipole response $|D_r|^2$ scaled by $r^2$. The peak in $r^2|D_r|^2$ marks the length scale $\xi$ above which the system behaves as a homogeneous elastic medium (vertical dashed lines). In Fig.~\ref{fig:length}(b) we show the mean square displacement $\langle \Delta R^2 \rangle$ of a random walk normalised to time $t$ for different $\phi$. For a long time, the signal plateau indicates a diffusive regime with coefficient $D_\infty$. We measured the crossover between the anomalous and diffusive regime with the criterion $\langle \Delta R^2 \rangle/t=1.15 D_\infty$, which gives us an estimate for $\xi$. Changing this threshold shifts the values for $\xi$ but not the scaling with $\phi$. In Fig.~\ref{fig:length}(c) we plot $\xi$ as a function of $\Delta\phi=\phi-\phi_c$ for both dipole responses and random walks. We find a very good agreement between the two approaches. Moreover, the divergence of $\xi$ approaching $\phi_c$ is consistent with our exponent $\nu\simeq1.3$ from the finite size study of the rigidity fraction discussed above. Comparing the dipole response results across different system sizes from $N=16$k to $N=128$k, we observe a significant finite-size effect close to the rigidity transition.

%%%%%%%%%%%%%%%%%%%%%%%%%%%%%%%%%%%%%%%%%%%%%%%%%%%%%%%%%%%%%%%%%%%%%%%
%%%%%%%%%%%%%%%%%%%%%%%%%%%%%%%%%%%%%%%%%%%%%%%%%%%%%%%%%%%%%%%%%%%%%%%

\subsection{Radial distributions and pore sizes}

Here we provide additional methods aimed at describing the architecture of our network gel and extracting a length scale for structural heterogeneities. First, we compute the integrated radial distribution function $N(r)$. As shown in Ref.~\cite{mizuno2021structural}, one can extract a length scale $\xi$ that marks the transition between the small scale fractal behaviour $N(r)\sim r^{d_f}$ and the far field scaling $N(r)\sim r^3$ (in 3D). Next, we measure an estimate of the pore size distribution associated with the percolating network. We use the method described in detail in Ref.~\cite{bhattacharya2006fast}, where a set of fictitious particles (labelled "ghost 1") is uniformly distributed throughout our sample. For each point, we measure the radius $R_{\rm max}$ of the largest spherical cavity that can be inserted without touching the spine. We then generate a second set of randomly inserted points (labelled "ghost 2"). Finally, we measure the largest void radius $R_{p}$ associated with the "ghost 1" set that encloses each point of the second "ghost 2" data set. We define the pore size as its diameter, namely $p_s=2R_{p}$.

\begin{figure*}[h!]
   \includegraphics[width = 1\textwidth]{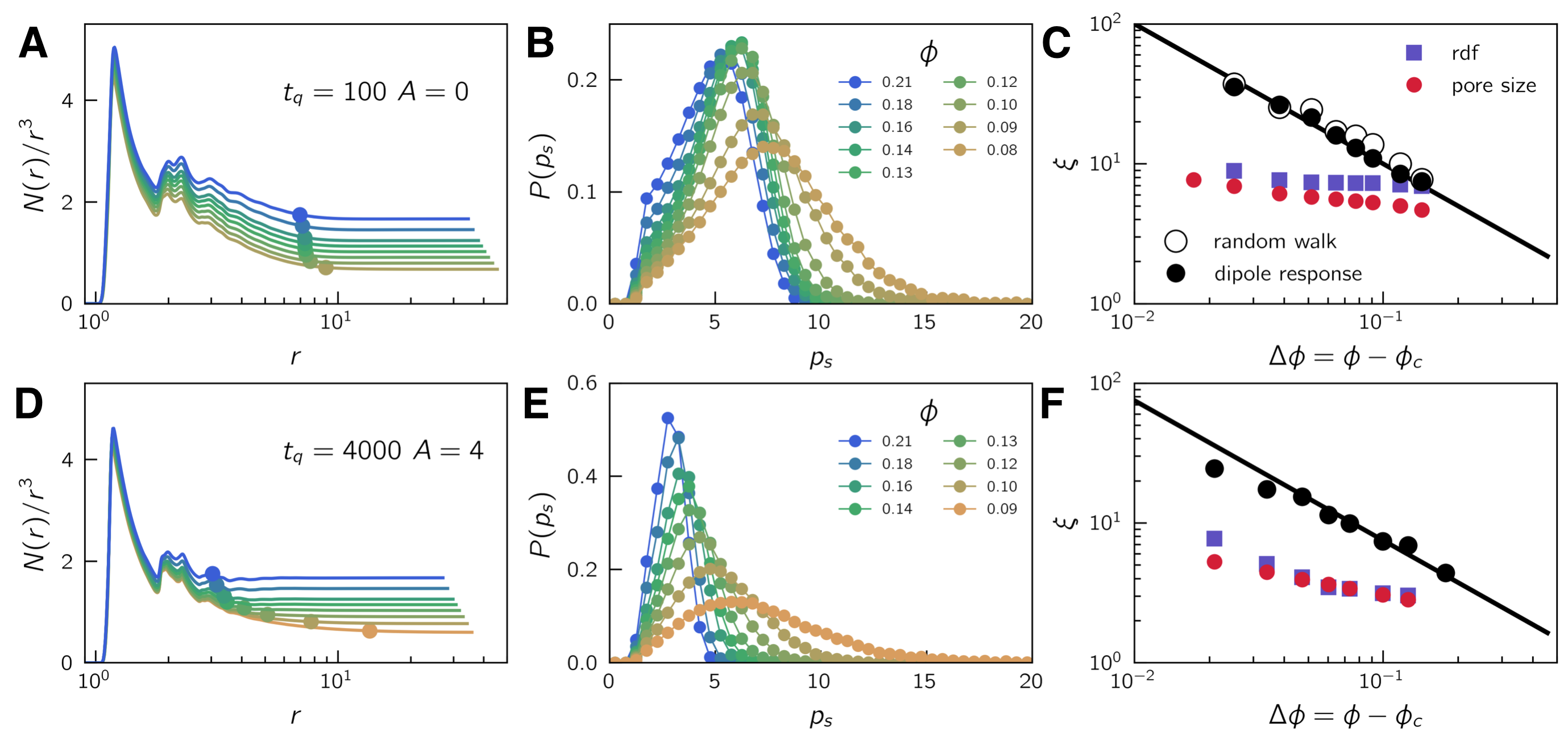}
  \caption{\footnotesize (a) Normalized integrated value of radial distribution function for gels formed with $t_q=100$ and $\kappa=0$. (b) Pore size distribution for different particle concentrations. (c) Length scale of heterogeneities plotted against the distance to the rigidity point for different methods. The solid line indicates $\xi=(\phi-\phi_c)^{-1}$. (d), (e), and (f) are the same as in (a), (b), and (c), respectively, but for $t_q=4000$ and $\kappa=4$. The solid in (f) indicates $\xi=0.75(\phi-\phi_c)^{-1}$.}
  \label{fig:rdf_pore}
\end{figure*}

In Fig.~\ref{fig:rdf_pore}(a) we show the normalised integrated radial distribution function $N(r)/r^3$ at different volume fractions $\phi$ for a rapidly quenched gel with $t_q=100$ and no electrostatic interaction $A=0$. We find only a small increase of $\xi$ with $\phi$ as shown in Fig.~\ref{fig:rdf_pore}(c). In Fig.~\ref{fig:rdf_pore}(b) we show the pore size distribution for the same densities. As $\phi$ decreases, the pore distribution widens with typical $p_s$ values shifting from $5$ to $10$ particle diameters. However, this growth is much weaker than the length scale scaling extracted from the dipole responses or the random walk as superimposed in Fig.~\ref{fig:rdf_pore}(c). Only the latter two have the consistent scaling with the critical exponent $\nu \simeq 1$ extracted from the finite size scaling. Note that we performed the same analysis on the gel formed by a longer quench time $t_q=4000$ and a large electrostatic repulsion $A=4$. The data are shown in Fig.~\ref{fig:rdf_pore}(d), (e) and (f). We arrive at the same conclusion, namely that the scaling of the typical pore size grows slower than the elastic length extracted from the linear response of the local strain. Furthermore, within the accuracy of our off-lattice simulations, it seems that the exponent controlling the divergence of the elastic length for the system self-assembling with the competitive interaction ($\kappa=4$) is the same as for the purely attractive case ($\kappa=0$), namely $\nu \simeq 1$. This result indicates that in 3D the zero frequency shear modulus follows $G_0\sim \xi^{-4}$.

%%%%%%%%%%%%%%%%%%%%%%%%%%%%%%%%%%%%%%%%%%%%%%%%%%%%%%%%%%%%%%%%%%%%%%%
%%%%%%%%%%%%%%%%%%%%%%%%%%%%%%%%%%%%%%%%%%%%%%%%%%%%%%%%%%%%%%%%%%%%%%%

\subsection{Fractal dimension}

\begin{figure*}[h!]
   \includegraphics[width = 1\textwidth]{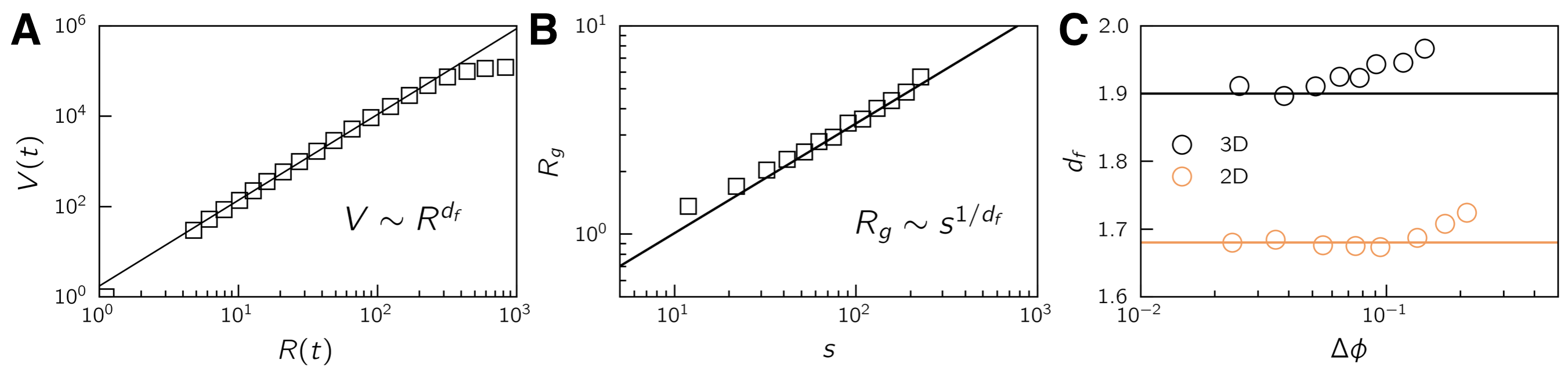}
  \caption{\footnotesize (a) Volume $V$ explored by the random walk as a function of the radius $R=\sqrt{\langle \Delta R^2\rangle}$. The solid line corresponds to the scaling $V\sim R^{d_f}$, with $d_f\simeq1.9$. (b) The gyration radius $R_g$ as a function of the cluster size $s$ measured close to $\phi_c$. The solid line indicates the scaling $R_g\sim s^{1/d_f}$, with $d_f\simeq1.9$. (c) The fractal dimension as a function of the distance to the rigidity point $\Delta\phi=\phi-\phi_c$ extracted from $V\sim R^{d_f}$ (circles) and $R_g\sim s^{1/d_f}$ (horizontal line) for both 2D and 3D network gels.}
  \label{fig:fractal_dimension}
\end{figure*}

The random walk on particle networks can also be used to extract the fractal dimension $d_f$ of our gels. In particular, we can plot the volume $V(t)$ explored by the tracer at time $t$ as a function of the radius $R(t)=\sqrt{\langle \Delta R^2\langle_t}$. Here $V$ is evaluated as the number of independent sites $N_s$ explored at a time $t$. For fractal objects we expect the relation $V\sim R^{d_f}$~\cite{nakayama1994dynamical}, see example in Fig.~\ref{fig:fractal_dimension}(a). In addition, we can calculate the scaling of the gyration radius $R_g$ with the cluster size $s$. Near the percolation point, $R_g$ scales as $s^{1/d_f}$~\cite{nakayama1994dynamical}, as shown in Fig.~\ref{fig:fractal_dimension}(b). In Fig.~\ref{fig:fractal_dimension}(c) we show $d_f$ as a function of $\Delta\phi$ for 2D and 3D gels. We find that the two methods agree, giving $d_f=1.68$ and $d_f=1.9$ for 2D and 3D samples respectively.

%%%%%%%%%%%%%%%%%%%%%%%%%%%%%%%%%%%%%%%%%%%%%%%%%%%%%%%%%%%%%%%%%%%%%%%
%%%%%%%%%%%%%%%%%%%%%%%%%%%%%%%%%%%%%%%%%%%%%%%%%%%%%%%%%%%%%%%%%%%%%%%

\subsection{Vibrational Density of States (VDoS)}

\begin{figure*}[h!]
   \includegraphics[width = 1\textwidth]{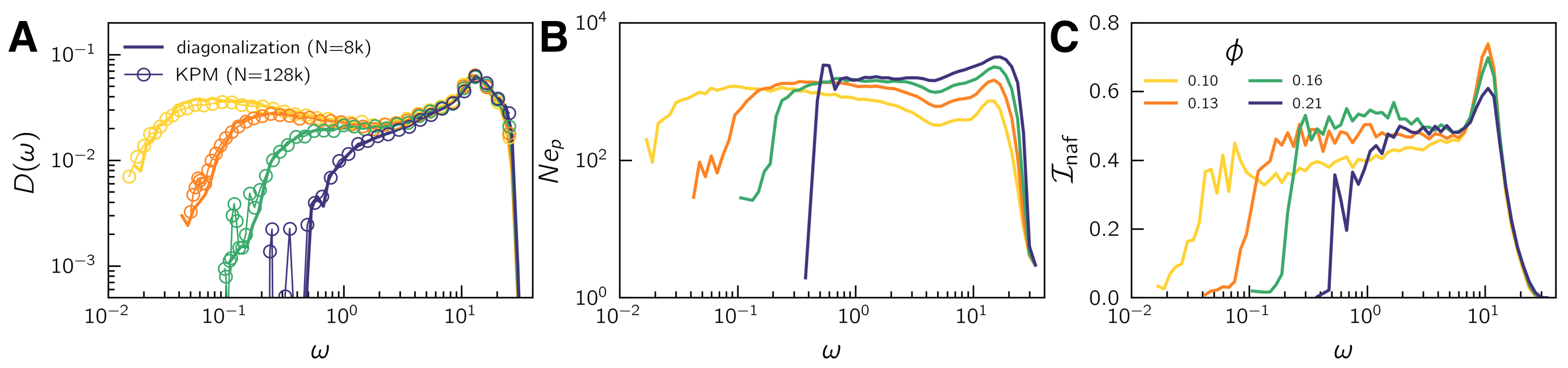}
  \caption{\footnotesize (a) Comparison of VDoS $D(\omega)$ obtained by exact diagonalisation (solid line) and KPM method (empty circles) at different concentrations. (b) Participation ratio $Ne_p$ versus mode frequency $\omega$. (c) Contribution of the mode to the non-affine shear modulus (see main text for details). Gels are formed with $\kappa=4$ and $t_{\rm quench}=4000$.}
  \label{fig:vdos_kpm}
\end{figure*}

In this section we provide supporting results on the vibrational properties of our gels. We show how to extract the (spectral) fracton dimension $\tilde{d}$ and the crossover frequency $\omega_c$. In Fig.~\ref{fig:vdos_kpm}(a) we compare the vibrational density of states (VDoS) extracted from an exact diagonalisation in small 3D systems ($N=8$k) and with the KPM method in a larger system ($N=128$k). We find perfect agreement between the two approaches. In Fig.~\ref{fig:vdos_kpm}(b) and (c) we show the mode participation ratio and the contribution to the non-affine shear modulus as a function of mode frequency, respectively. We find that the modes populating the VDoS plateau are relatively spatially extended and have a good coupling with the shear force $\Theta$, leading to a reduction in the bulk shear modulus. As in decompressed packings close to the jamming transition, a plateau of soft modes with good coupling to the loading geometry emerges, giving rise to a power-law regime in the viscoelastic spectrum.

%%%%%%%%%%%%%%%%%%%%%%%%%%%%%%%%%%%%%%%%%%%%%%%%%%%%%%%%%%%%%%%%%%%%%%%
%%%%%%%%%%%%%%%%%%%%%%%%%%%%%%%%%%%%%%%%%%%%%%%%%%%%%%%%%%%%%%%%%%%%%%%

\begin{figure*}[h!]
   \includegraphics[width = 1\textwidth]{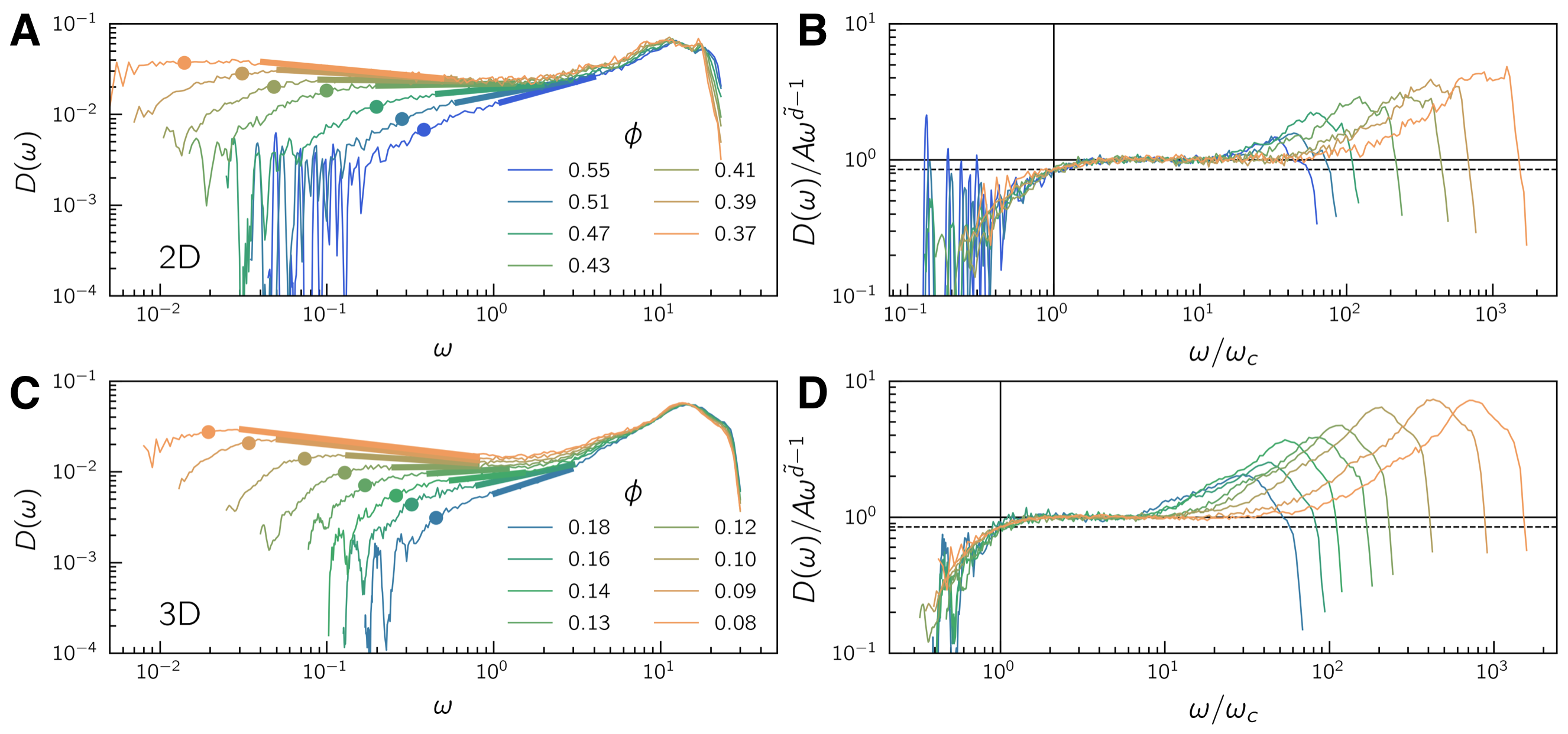}
  \caption{\footnotesize (a) VDoS $D(\omega)$ for two-dimensional gels at different volume fractions. The thick solid lines indicate the linear fit to extract the fracton dimension with $D(\omega)=A\omega^{\tilde{d}-1}$. (b) Collapsed VDoS using the fracton dimension $\tilde{d}$ and crossover frequency $\omega_c$. The norm is obtained from the same linear fit of the VDoS plateau and $\omega_c$ is extracted with the criterion $D(\omega)/(A\omega^{\tilde{d}-1}) < 0.85$ (horizontal dashed line). (c) and (d) are the same as (a) and (b) but for three-dimensional network gels. Gels are formed using $\kappa=0$ and $t_{\rm quench}=100$.}
  \label{fig:vdos_collapse}
\end{figure*}

In Fig.~\ref{fig:vdos_collapse}(a) we show the VDoS of two-dimensional gels at different concentrations. We extract the fracton dimension by fitting the plateau of $D(\omega)$ by $A\omega^{\tilde{d}-1}$, as indicated by the thick solid coloured lines. The crossover frequency is estimated with the criterion $D(\omega)/(A\omega^{\tilde{d}-1}) < 0.85$ (horizontal dashed line in Fig.~\ref{fig:vdos_collapse}(b) and (d)). In Fig.~\ref{fig:vdos_collapse}(b) we show the resulting low frequency collapse of all VDoS with different volume fractions. Results for 3D gels are shown in Fig.~\ref{fig:vdos_collapse}(c) and (d).

%%%%%%%%%%%%%%%%%%%%%%%%%%%%%%%%%%%%%%%%%%%%%%%%%%%%%%%%%%%%%%%%%%%%%%%
%%%%%%%%%%%%%%%%%%%%%%%%%%%%%%%%%%%%%%%%%%%%%%%%%%%%%%%%%%%%%%%%%%%%%%%

\end{document}